\documentclass[conference]{IEEEtran}
\ifCLASSINFOpdf
  \usepackage[pdftex]{graphicx}
  \DeclareGraphicsExtensions{.pdf,.jpeg,.png}
\else
\fi
\usepackage[cmex10]{amsmath}
\usepackage[tight,footnotesize]{subfigure}
\usepackage{url}

\begin{document}
%
\title{High Performance Computing of Gene Regulatory Networks using a Message-Passing Model}

\author{\IEEEauthorblockN{Kimberly Glass}
\IEEEauthorblockA{Brigham and Women's Hospital and\\Harvard Medical School\\
Boston, MA, USA\\
Email: rekrg@channing.harvard.edu}
\and
\IEEEauthorblockN{John Quackenbush}
\IEEEauthorblockA{Dana-Farber Cancer Institute and\\Harvard School of Public Health\\
Boston, MA, USA\\
Email: johnq@jimmy.harvard.edu}
\and
\IEEEauthorblockN{Jeremy Kepner}
\IEEEauthorblockA{Lincoln Laboratory and \\ Massachusetts Institute of Technology\\
Cambridge, MA, USA\\
Email: kepner@ll.mit.edu}}


%


\maketitle

\begin{abstract}
Gene regulatory network reconstruction is a fundamental problem in computational biology. We recently developed an algorithm, called PANDA (Passing Attributes Between Networks for Data Assimilation), that integrates multiple sources of 'omics data and estimates regulatory network models. This approach was initially implemented in the C++ programming language and has since been applied to a number of biological systems. In our current research we are beginning to expand the algorithm to incorporate larger and most diverse data-sets, to reconstruct networks that contain increasing numbers of elements, and to build not only single network models, but sets of networks. In order to accomplish these ``Big Data'' applications, it has become critical that we increase the computational efficiency of the PANDA implementation. In this paper we show how to recast PANDA's similarity equations as matrix operations. This allows us to implement a highly readable version of the algorithm using the MATLAB/Octave programming language. We find that the resulting M-code much shorter (103 compared to 1128 lines) and more easily modifiable for potential future applications. The new implementation also runs significantly faster, with increasing efficiency as the network models increase in size. Tests comparing the C-code and M-code versions of PANDA demonstrate that this speed-up is on the order of 20-80 times faster for networks of similar dimensions to those we find in current biological applications.
\end{abstract}


%
\IEEEpeerreviewmaketitle

\section{Introduction}

Rapidly evolving genomic technologies are providing biologically informative data of unprecedented volume, velocity, and variety, with the potential to yield new insights into the processes driving disease. This data has allowed us to develop a more unified understanding of how many different types of interactions at multiple and vastly different scales can influence biological systems. We now appreciate that changes in cellular states involve simultaneous alterations to the genome, epigenome, transcriptome, metabolome, and proteome of the cell. These are often characterized by complex networks whose structures are altered as the phenotype changes. The generation of Big Data in this area now presents an unprecedented opportunity to develop scalable methods capable of modeling these networks, informing us about the nature of disease and ultimately allowing us to hypothesize about the therapeutic approaches most appropriate for each disease state.

Working in this area, we recently developed and published a method, called PANDA (Passing Attributes between Networks for Data Assimilation), that uses a ``message passing'' approach to integrate multiple types of genomic data and construct directed genome-wide regulatory networks \cite{PANDA}. PANDA models network interactions as communication between ``transmitters'' and ``receivers'', referred to in the original publication as ``effector'' and ``affected'' nodes. This approach recognizes that for communication to occur, both the transmitter and the receiver have an essential role. By constructing a ``prior'' regulatory network consisting of potential routes for communication and integrating with other sources of information, PANDA estimates the \emph{responsibility} and \emph{availability} of each potential interaction, predicts where communication is succeeding or failing, and deduces condition-specific network structures. While many methods exist for inferring relationships in a gene regulatory network \cite{DeSmetMarchal,DREAM5}, PANDA represents the first method that incorporates multiple biological data types naturally, by comparing them in order to emphasize common elements. Most importantly, this approach provides a unified modeling framework for integrating multiple and heterogeneous biological data types.

We have now applied PANDA to explore the effects of smoking in knockout mouse models \cite{HHIPPANDA}, to search for potential drug candidates in ovarian cancer \cite{OvCaPANDA} and to build sex-specific regulatory networks in chronic obstructive pulmonary disease (COPD) \cite{SDPANDA}. In each of these applications we compared pairs or sets of networks reconstructed using PANDA to uncover regulatory mechanisms that would not have been identified using gene-based approaches. The importance of comparing network states highlights the need for a computationally efficient implementation of the PANDA algorithm. Indeed, much of our ongoing and future research now revolves around computing and comparing multiple versions and various sets of these already-large network models.

The PANDA algorithm for network reconstruction was originally implemented in the C++ programming language. In order to increase code readability (to facilitate future modification and data integration) and to enable greater scaling of the algorithm, which will be necessary in order to incorporate larger and increasingly diverse data-sets, we have implemented a version of the algorithm in the MATLAB/Octave programming language. We find that this M-code implementation of PANDA runs significantly faster than the C++ version across a range of network sizes. The M-code is also much shorter (103 versus 1128 lines). In this paper we outline the approach we took in re-coding the algorithm. Namely, we begin by reviewing the mathematical framework used by PANDA. We then show how PANDA's similarity equation can be re-written as products of matrices, enabling us to take advantage of MATLAB's extreme optimization of these operations. Finally, we systematically compare the computational efficiency of the previous and new implementations of the algorithm using a range of input data sizes, including those we often encounter in real-world biological applications.

\section{Approach}

\subsection{Finding Agreement between Networks using PANDA}

\begin{figure}[!t]
\centering
\includegraphics[width=3.5in]{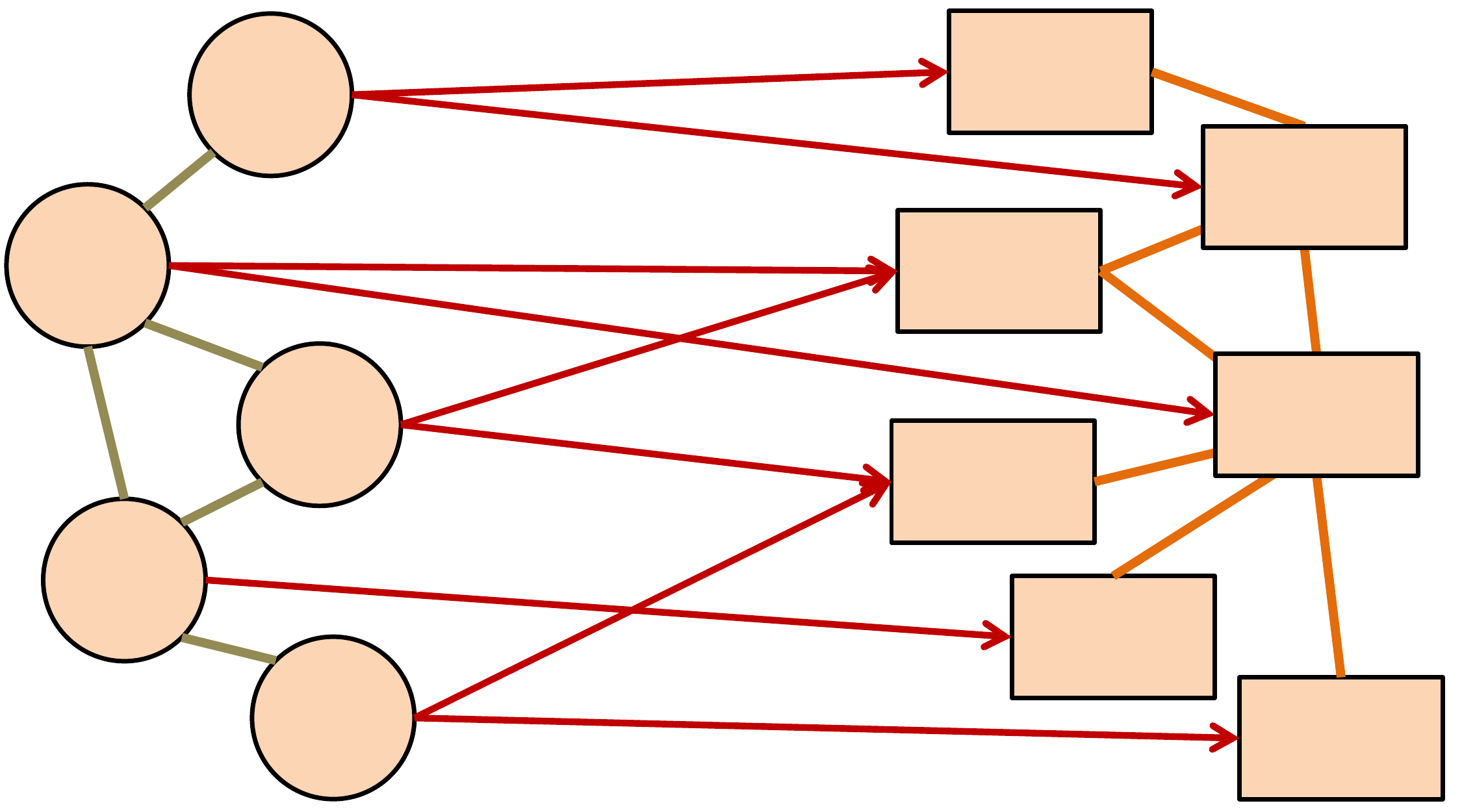}
\caption{The PANDA network reconstruction algorithm models two types of nodes: ``effectors'' (circles) and ``affected'' (squares). It also considers three types of networks: ``cooperativity'' (between pairs of ``effector'' nodes, brown lines), ``regulatory'' (from ``effector'' to ``affected'' nodes, red lines), and ``co-regulatory'' (between pairs of ``affected'' nodes).}
\label{pandanets}
\end{figure}

Transcriptional regulation depends on a complex relationship between transcription factors (TFs) and their downstream targets. To model this process, we developed PANDA, a ``message-passing'' algorithm for reconstruction of gene regulatory networks. PANDA considers two types of network nodes, ``effectors'' and ``affected'', and models three types of network relationships between these nodes (see Figure \ref{pandanets}). In the context of biological regulatory networks, transcription factors can be viewed as effector nodes that regulate (affect) the behavior of their downstream target genes.

\begin{figure}[!t]
\centering
\includegraphics[width=3.5in]{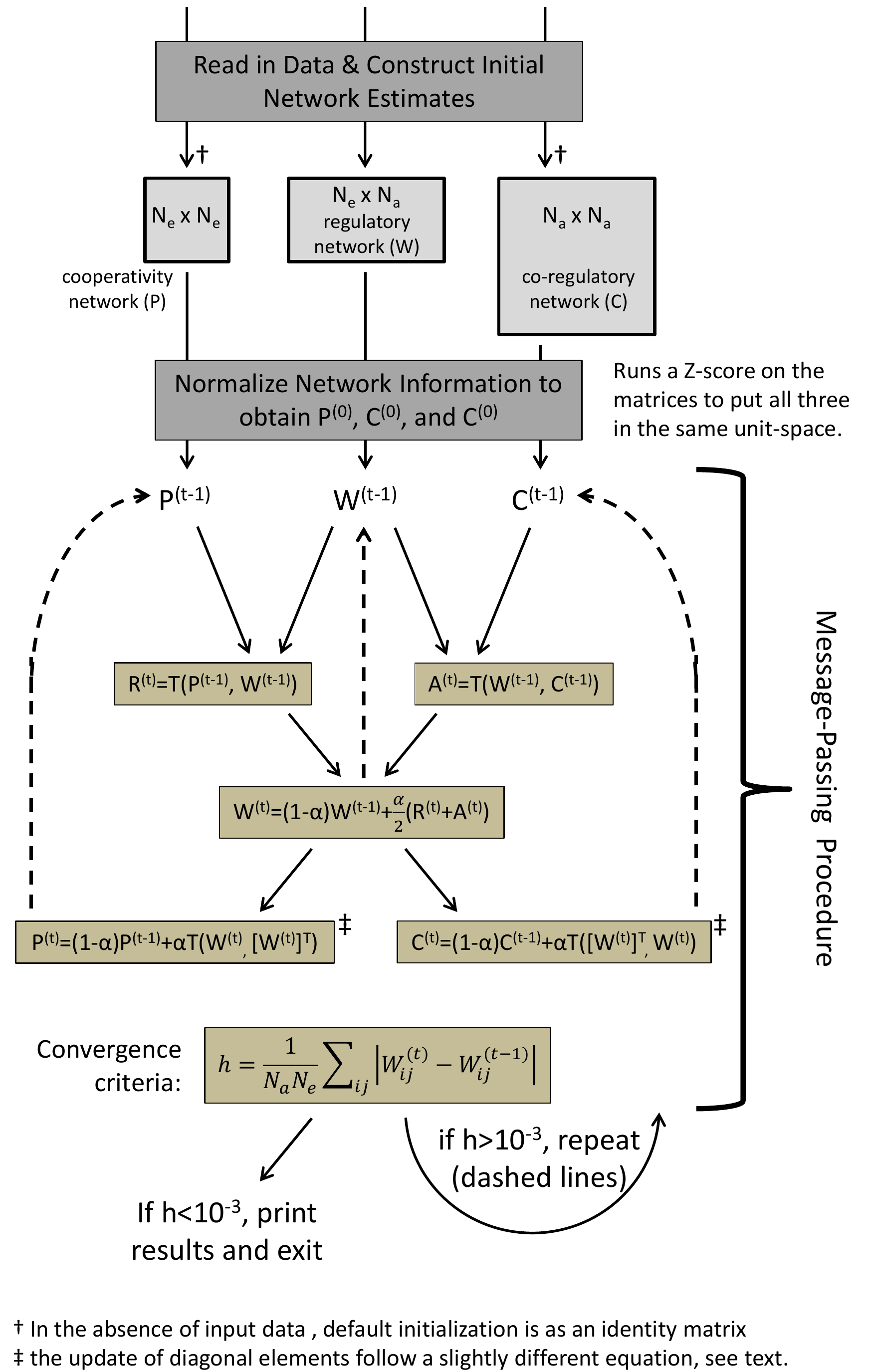}
\caption{Overview of the steps included in the PANDA network reconstruction algorithm. A significant portion of the algorithm involves a message-passing procedure in which information from the initial data is slowly merged together.}
\label{PANDA_Algorithm}
\end{figure}

An overview of the PANDA algorithm is presented in Figure \ref{PANDA_Algorithm}. To begin, PANDA takes input information pertaining to the relationships between effector and affected nodes and constructs three ``seed'' networks. A symmetric network between effector nodes (of dimensions $N_e \times N_e$, where $N_e$ is the number of effector nodes) is referred to as the ``cooperativity'' network ($P$). Similarly, a symmetric network between affected nodes (of dimensions $N_a \times N_a$, where $N_a$ is the number of affected nodes) is referred to as the ``co-regulatory'' network ($C$). Finally, a non-symmetric network from effector to affected nodes (of dimensions $N_e \times N_a$) is called the ``regulatory'' network ($W$).

In the context of transcriptional data, PANDA reads in expression information and estimates an initial co-regulatory network by calculating the Pearson correlation between pairs of genes across all the samples in the data. The cooperativity network is initially defined based on a user-provided set of interactions between pairs of transcription factors. To address the incompleteness of biological data, in the absence of expression or protein interaction information PANDA will initialize these two networks as identity matrices. Finally, to create the initial regulatory network, PANDA reads in predicted transcription factor-gene relationships, which are often estimated by using DNA sequence information to create a edge between a transcription factor (effector node) and a target gene (affected node) if a known binding ``motif'' \cite{DNAMotif} for that transcription factor exists in the promoter region of the gene. In order to effectively combine information from these diverse data sources, PANDA performs a Z-score normalization on each of the networks such that the edge-weights contained in their corresponding matrices all are in the same unit-space and of similar distribution.

After reading in and normalizing the input data, PANDA performs a message-passing procedure to slowly integrate the information contained in the three initial networks. One important aspect of this approach is its emphasis on agreement among network neighborhoods rather than direct targeting information. For example, unlike many other network reconstruction approaches \cite{DeSmetMarchal,DREAM5}, PANDA does not infer a relationship between a transcription factor and a gene directly from that pair's co-expression; instead, an edge is inferred if the gene is co-expressed with \emph{other} targets of the transcription factor. By performing iterative ``soft'' updates on each of the networks, the models slowly accumulate evidence for interactions that is shared across all of the input data sources, eventually moving to consensus networks that better explain the full set of observations. This method thus serves to infer comprehensive new biology that would not be obvious based on any single data-type alone.

Supporting this idea, at the heart of PANDA is an equation that evaluates the similarity between sets of interactions in two networks:
\begin{equation}
\begin{split}
T_{xy} & = T(\vec{x}, \vec{y}) \\
& =\frac{\vec{x}\cdot\vec{y}}{\sqrt{\|\vec{x}\|^2+\|\vec{y}\|^2-|\vec{x}\cdot\vec{y}|}}\\
& =\frac{\sum_k {x_k y_k}}{\sqrt{\sum_k x_k^2+\sum_k y_k^2-|\sum_k x_k y_k|}}
\end{split}
\label{Tanimoto}
\end{equation}
This equation is used repeatedly throughout the PANDA message-passing procedure. Below we detail how this equation has been and can be implemented to best take advantage of the strengths of different coding languages. We show that changing how we formulate this and the other mathematics in PANDA can have a dramatic impact in the implementation and can drastically speed up algorithm.

\subsection{The PANDA Message-Passing Algorithm}

The message-passing framework of PANDA is designed to slowly merge information from different network structures, initially derived from an underlying set of biological data. In order to do this, PANDA iteratively updates each of three initial networks. Namely, at each iteration, $t$, PANDA first updates the regulatory network based on the \emph{responsibility} ($R$) and \emph{availability} ($A$) of edges:
\begin{equation}
W_{ij}^{(t)}=(1-\alpha)W_{ij}^{(t-1)}+\frac{\alpha}{2}(A_{ij}^{(t)}+R_{ij}^{(t)})
\label{Wupdate}
\end{equation}
The \emph{responsibility} of an edge between TF $i$ and gene $j$ in the regulatory network is calculated based the previous level of agreement between the set of transcription factors that target gene $j$ ($W_{.j}^{(t-1)}$) and those that cooperate with transcription factor $i$ ($P_{i.}^{(t-1)}$):
\begin{equation}
R_{ij}^{(t)}=T(P_{i.}^{(t-1)}, W_{.j}^{(t-1)})
\label{Req}
\end{equation}
Similarly, the \emph{availability} of each edge is based on the level of agreement between the regulatory targets of transcription factor $i$ ($W_{i.}^{(t)}$) and the set of genes with which gene $j$ is co-regulated ($C_{.j}^{(t)}$):
\begin{equation}
A_{ij}^{(t)}=T(W_{i.}^{(t-1)}, C_{.j}^{(t-1)})
\label{Aeq}
\end{equation}

Next PANDA updates the protein cooperativity and gene co-regulatory networks by comparing the sets of genes regulated by a pair of TFs, and the sets of TFs targeting a pair of genes, respectively. Namely:
\begin{equation}
\begin{split}
P_{i \neq m}^{(t)} & = (1- \alpha) P_{i \neq m}^{(t-1)} + \alpha T(W_{i.}^{(t)}, [W_{p.}^{(t)}]^T) \\
C_{k \neq j}^{(t)} & = (1- \alpha) C_{k \neq j}^{(t-1)} + \alpha T([W_{.k}^{(t)}]^T, W_{.j}^{(t)})
\end{split}
\label{PCupdate}
\end{equation}
To satisfy convergence criteria, the diagonal elements of $P$ and $C$ are updated based on the off-diagonal elements of $P$ and $C$, respectively. More specifically:
\begin{equation}
\begin{split}
P_{ii}^{(t)} & = (1 - \alpha) P_{ii}^{(t-1)} + N_a \sigma_{i}^{(t)} e^{2 \alpha t} \\
C_{jj}^{(t)} & = (1 - \alpha) C_{jj}^{(t-1)} + N_e \sigma_{j}^{(t)} e^{2 \alpha t}
\end{split}
\label{PCdiagonal}
\end{equation}
where $\sigma_i$ is the standard deviation across the non-diagonal elements in row  $i$ of $P$ ($P_{i.}^{(t)}$), and $\sigma_j$ is the standard deviation across the non-diagonal elements in row $j$ of $C$ ($C_{j.}^{(t)}$).

Finally, at each iteration, when computing $W^{(t)}$, PANDA also calculates the hamming distance between the previous and predicted regulatory network:
\begin{equation}
h=\frac{1}{N_e N_a}\sum_{ij} |W_{ij}^{(t)}-W_{ij}^{(t-1)}|
\label{Hamming}
\end{equation}
The updates (of $W$, then $P$ and $C$) are iteratively repeated until convergence, which is defined as $h < 10^{-3}$. Depending on the update parameter, $\alpha$, this can involve anywhere from ten to several hundred iterations of the message-passing procedure; the current suggested default value for $\alpha$ is $0.1$ and involves approximately $40$ iterations. The result of the PANDA message-passing approach is a value associated with each of the edges in all three networks that reflects shared structure across the original set of input data.

\subsection{Implementing PANDA using Matrix Algebra}

$T$ (Equation \ref{Tanimoto}) is used to calculate the values for each edge in $A$, $R$, $C$ and $P$ (see Equations \ref{Req}-\ref{PCupdate}). As it is written, Equation \ref{Tanimoto}  returns a single number, namely, a score for an edge from node $i$ to node $j$ in one network, based on vectors containing values from the $i^{th}$ row and $j^{th}$ column of matrix representations for two other networks. In the C++ implementation, this is calculated via summations within nested for-loops. Below we show an example of the general framework for performing this operation, based on predicting the \emph{availability} ($A$, equation \ref{Aeq}) for all the edges. Note that this is not exactly replicated from the C++ code (available at \url{http://sourceforge.net/projects/panda-net/}), but does accurately represent the way the similarity calculation is implemented:
\begin{equation}
\begin{split}
& for(i=0; i<Ne; i++)\{ \\
& \ \ \ \ for(j=0; j<Na; j++)\{ \\
& \ \ \ \ \ \ \ \ Avar=0; \ Bvar=0; \ Cvar=0; \\
& \ \ \ \ \ \ \ \ for(k=0; k<Ne; k++)\{ \\
& \ \ \ \ \ \ \ \ \ \ \ \ Avar+=C[j].tar[k]*W[i].tar[k]; \\
& \ \ \ \ \ \ \ \ \ \ \ \ Bvar+=C[j].tar[k]*C[j].tar[k]; \\
& \ \ \ \ \ \ \ \ \ \ \ \ Cvar+=W[i].tar[k]*W[i].tar[k]; \\
& \ \ \ \ \ \ \ \ \ \} \\
& \ \ \ \ \ \ \ \ A[i].tar[j]=Avar/sqrt(Bvar+Cvar-fabs(Avar)); \\
& \ \ \ \ \} \\
& \}
\end{split}
\label{Ccode}
\end{equation}
Here we point out that instead of solving for each element of the $T$ matrix individually, as it done in the code-snippet shown above, one can alternately solve for the entire matrix of T-values. Using the MATLAB/Octave programming language, this can be written as a series of matrix operations contained in the following 5 lines of code:
\begin{equation}
\begin{split}
& function \ Amat=Tfunction(X,Y); \\
& Amat=(X*Y); \\
& Bmat=repmat(sum(Y.^{\wedge}2,1), size(X,1), 1); \\
& Cmat=repmat(sum(X.^{\wedge}2,2), 1, size(Y,2)); \\
& Amat=Amat./sqrt(Bmat+Cmat-abs(Amat));
\end{split}
\label{Mcode}
\end{equation}
This matrix-oriented implementation requires the simultaneous storage of three potentially very large matrices ($Amat$, $Bmat$ and $Cmat$) in addition to the original given networks ($X$ and $Y$). On the other hand, the C++ implementation shown above, by estimating each element of T separately, only requires storing one of these matrices ($A$). Thus by solving for all the elements of the matrix at once we have increased the memory requirements for this process.

We have evaluated how to represent all of PANDA's mathematics, including the calculation of $T$, as matrix operations and re-implemented the algorithm in the MATLAB/Octave programming language (hereafter referred to as the M-code). This PANDA implementation includes five files:
\begin{enumerate}
\item \emph{RunPANDA.m}: The master M-file that reads in the input data, constructs the initial networks, runs the message-passing algorithm and prints the final network estimates to a file. This code calls two functions contained in other files: \emph{NormalizeNetwork()} and \emph{PANDA()}.
\item \emph{NormalizeNetwork.m}: Performs a Z-score transformation on a given input network.
\item \emph{PANDA.m}: Given set of three networks: $W$, $C$ and $P$, performs the core message-passing procedure. This code calls two functions contained in other files: \emph{Tfunction()} and \emph{UpdateDiagonal()}.
\item \emph{Tfunction.m}: Computes the similarity between two networks (Equation \ref{Tanimoto}).
\item \emph{UpdateDiagonal.m}: Updates the diagonal elements of a given symmetric network (Equation \ref{PCdiagonal}).
\end{enumerate}
The M-code can be downloaded at: \url{https://sites.google.com/a/channing.harvard.edu/kimberlyglass/tools/panda}.

\subsection{Comparing the M-code and C-code PANDA Implementations}

We have used \emph{cloc} to evaluate the number of lines of code in this new M-code implementation and compared that to the number of lines of code in the original C-code implementation of PANDA. We find that the M-code version of PANDA is approximately ten-times shorter (103 lines) than the the C-code (1128 lines). A small portion of this reduction in line-count can be attributed to additional features found in the C-code that we did not fully implement in the M-code. However, it is unlikely that adding these to the M-code would increase the line-count significantly. For example, the C-code includes a parameter that allows the user to randomize node-labels upon input; such an addition to the M-code would require only a few lines. In addition to being shorter, it is also important to emphasize that the new M-code implementation of PANDA is also highly human-readable, especially in comparison to the C-code. Thus integrating any ``missing'' features into the M-code, such as this randomization procedure, will be less cumbersome than the future introduction of other new features into the C-code.

In addition to code-length, we have also tested the speed of the C-code and M-code implementations of PANDA. In order to do this, in the C-code we added a call to the \emph{clock()} function immediately prior to and immediately after the message-passing procedure. Time in seconds was then determined by multiplying the difference in these calls by $CLOCKS\_PER\_SEC$. For the M-code, the run-time of the message-passing procedure was determined by the \emph{tic/toc} function, again placed immediately prior to and past the message-passing procedure. By placing \emph{clock()} and \emph{tic/toc} around the message-passing procedure, we are not addressing any single time-costs that are associated with reading in or parsing the original data, or in outputting the results.

For this speed-test we constructed one hundred random initial regulatory networks ($W$) at each of several various sizes ($N_e=N_a=\{125,250,500,1000,2000\}$). For the other two networks ($P$ and $C$) we took advantage of PANDA's default behavior wherein those are initialized as identity matrices in the lack of external information (see Figure \ref{PANDA_Algorithm}). We compiled the C-code using g++ with an optimization flag ($-O3$). We ran the M-code using both GNU Octave (version 3.8.2) and MATLAB R2014a (8.3.0.532) with the \emph{--singleCompThread} flag (see below for more information). All analyses were run on the MIT SuperCloud which uses the x86\_64 GNU/Linux operating system \cite{Reuther_llsupercloud:sharing}.

\begin{table}[!b]
\renewcommand{\arraystretch}{1.3}
\centering
\begin{tabular}{| c | c | c | c |}
  \hline
  $N_e=N_a$ & C-code & Octave & MATLAB \\
  \hline
  125 & $1.05 \pm 0.01$ & $0.50 \pm 0.03$ & $0.13 \pm 0.04$ \\
  250 & $4.65 \pm 0.03$ & $0.62 \pm 0.01$ & $0.54 \pm 0.03$ \\
  500 & $32.73 \pm 0.07$ & $2.82 \pm 0.03$ & $2.61 \pm 0.002$ \\
  1000 & $241.51 \pm 0.12$ & $15.82 \pm 0.09$ & $15.06 \pm 0.04$ \\
  2000 & $1954.5 \pm 33.12$ & $98.59 \pm 0.89$ & $95.04 \pm 0.25$ \\
  \hline
\end{tabular}
\vspace{2mm}
\caption{A comparison of the average run-time, in seconds, of the C-code and M-code implementations of the PANDA algorithm across 100 random input regulatory network models. The M-code was run using both Octave and MATLAB.}
\label{TimesTable}
\end{table}

The average and standard deviation in run-time across the one-hundred tests for each primary network size is shown in Table \ref{TimesTable}. We find the MATLAB and Octave runs of PANDA have comparable run-times and both are significantly faster than the C-code implementation. Even more importantly, the fold-improvement in speed in the MATLAB and Octave runs increases as the network sizes increase.

\subsection{Taking Advantage MATLAB's Built-in Multi-Threading Capabilities}

In order to more fairly compare results between the compiled C-code and the M-code runs of PANDA in both MATLAB and Octave, the results presented in Table \ref{TimesTable} ran the M-code implementation of PANDA after opening MATLAB using the \emph{--singleCompThread} option. However, by default, MATLAB has the ability to take advantage of the multi-threading capabilities of the computer on which it is installed. Therefore, next we determined if an even greater speed-up might be obtained by taking advantage of this multi-threading capability. To test this, we used the approach described above and constructed one single random input network for each of a range of sizes. We then evaluated the time it took to perform the message-passing procedure using the compiled C-code and the M-code when running MATLAB using either the single-thread or multi-threading (default) option.

\begin{figure}[!t]
\centering
\subfigure[The C-code and single- and multi-thread M-code Run-times]{\includegraphics[width=3.5in]{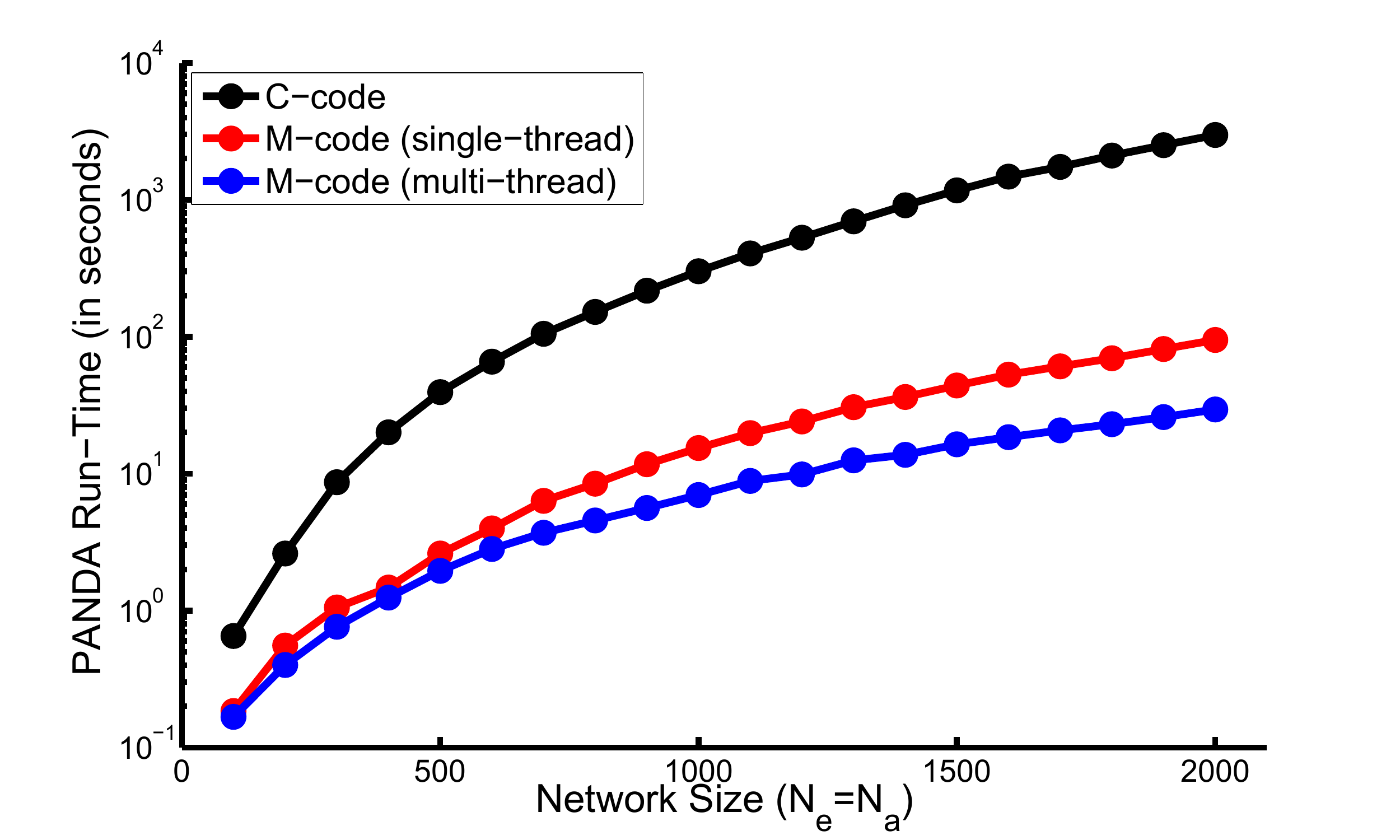}\label{DirectCompare}}
\subfigure[Ratio of the C-code versus the single- and multi-thread M-code run-times]{\includegraphics[width=3.5in]{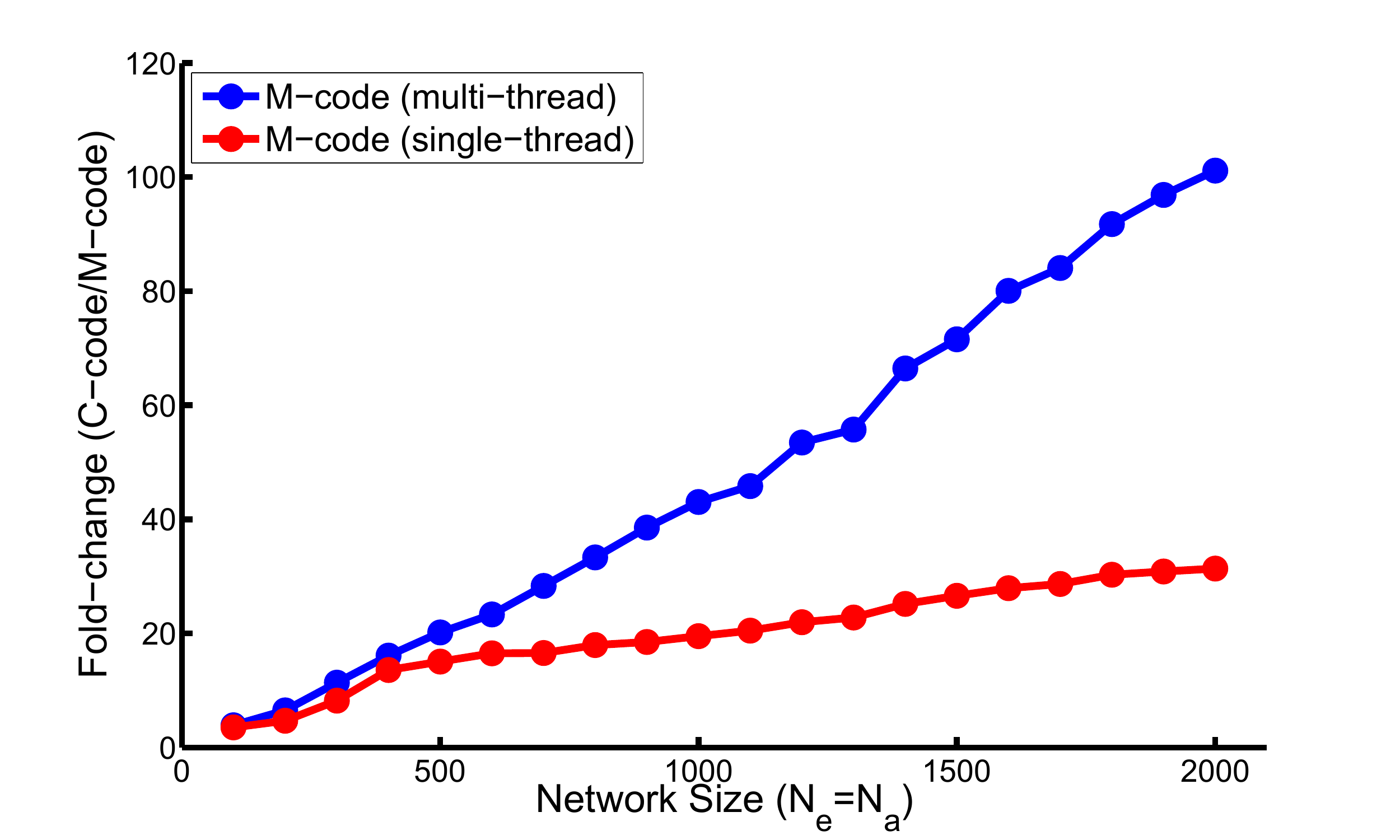}\label{RatioCompare}}
\caption{Comparison of times when running the M-code implementation of PANDA from within MATLAB using either the default multi-threading capabilities, or using only a single computational thread.}
\label{SingleVsMultiThread}
\end{figure}

The results of this test are plotted in Figure \ref{SingleVsMultiThread}. We find that by taking advantage of MATLAB's multi-threading capabilities we are able to decrease the run-time of the M-code significantly, and that this improvement continues to increase as the size of the network grows. In fact, for networks with several thousand nodes, the single-thread M-code improvement compared to the C-code is about 30-fold, but the multi-thread M-code run-time improvement compared to the C-code is over 100-fold. Although the magnitude of this speed-up will be dependent on hardware of the computer system on which MATLAB is installed, the real-life implications are profound. A network that would previously have taken several days to reconstruct using the C-code implementation of PANDA could be run in a matter of hours on a computer system with multi-threading capabilities. So far our tests of the message-passing procedure have not considered networks that take more than a few hours to reconstruct, even using the C-code. As we explain below, large networks necessitating a very long computation time are often encountered in the field of computational biology.

\subsection{Real-world Implications for Faster Network Reconstruction}

\begin{figure*}[!t]
\centering
\includegraphics[width=7in]{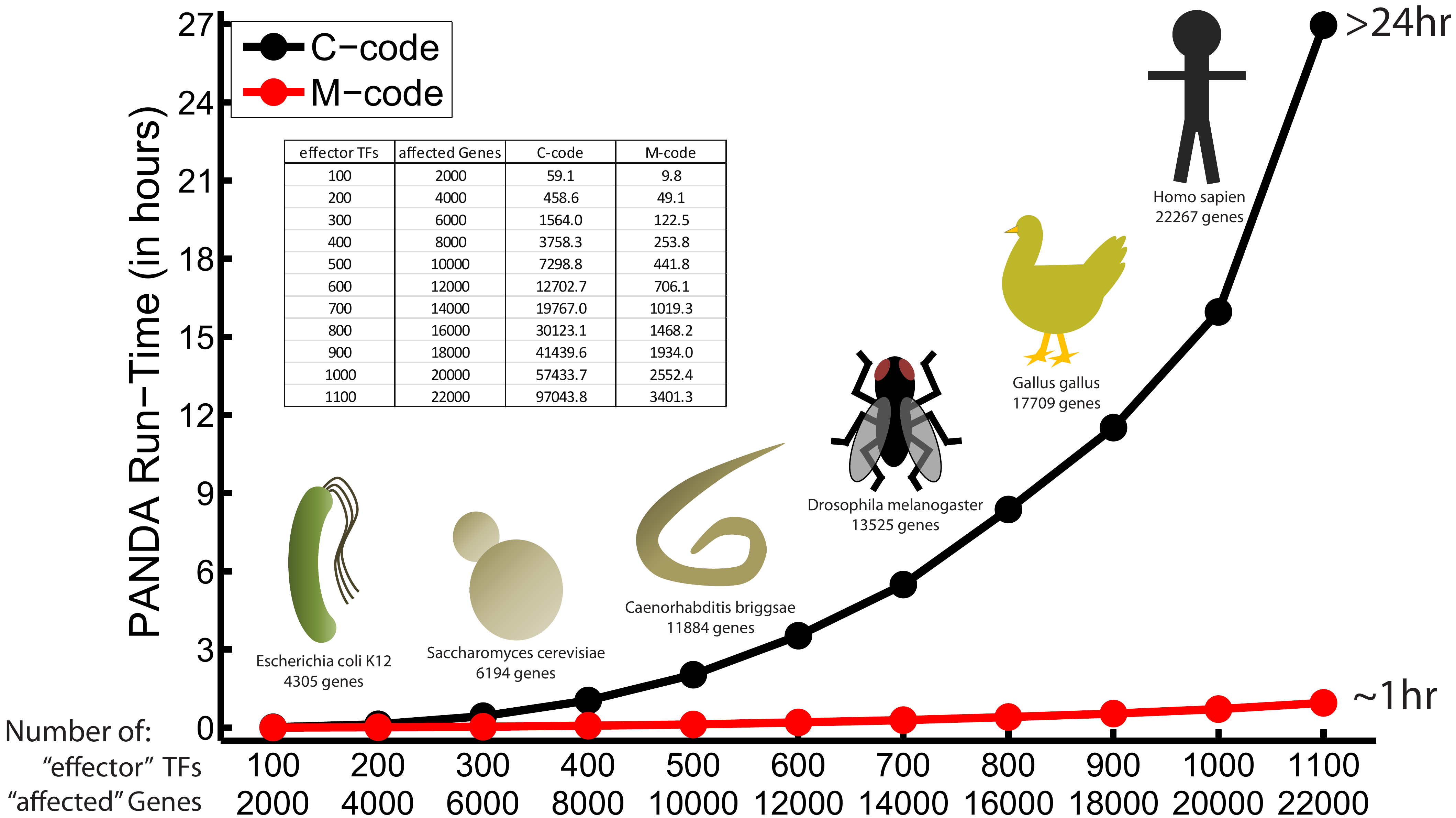}
\caption{An evaluate of the time PANDA needs to reconstruct networks that are of similar size to those found in real biological systems. Results from the tests in seconds are included in the table.}
\label{RealTime}
\end{figure*}

In biological systems, networks are often on the order of thousands, rather than hundreds of nodes. Even the genomes of ``smaller'' organisms contain thousands of genes (approximate six thousand for yeast and thirteen thousand for fruit fly). Current research on the human genome puts the number of protein-coding genes at approximately twenty to twenty-five thousand. Fortunately, the gene regulatory networks for biological systems are often highly a-symmetric with respect to their ``effector'' and ``affected'' nodes, with the number of transcription factor regulators (``effector'' nodes) typically only 5-10\% that of genes. With these real-world values in mind, we tested the speed of our C-code and M-code implementations of PANDA across a set of ``realistic'' network sizes. Here we again ran MATLAB with the \emph{--singleCompThread} option for a more fair comparison with the C-code. The results are shown in Figure \ref{RealTime}.

Using the initial C-code implementation of PANDA, a ``human-sized'' network reconstruction took over a day. However, using the new M-code implementation of PANDA this same analysis was accomplished in approximately one hour. It is also important to emphasize that if we were to take advantage of MATLAB's default multi-threading capabilities this time would be reduced even further. A ``real-world'' network that might take approximately twenty-four hours to compute using the C-code could possibly be computed in the matter of minutes using the M-code. Overall this has significant implications both for biomedical research as well as for our ability to provide real-time results as we are beginning to integrate network analysis into clinical applications.

\section{Conclusion}

The original implementation of the PANDA algorithm was coded in C++ and determined shared information between networks based on a similarity equation implemented through a series of nested for-loops. Here we showed how to re-cast that equation as a series of matrix operations in order to re-code PANDA in the MATLAB/Octave programming language. We find that this  ``M-code'' implementation not only greatly increases code readability compared to the ``C-code'', it also drastically \emph{decreases} the time needed to compute the networks. Importantly, the M-code implementation also has the potential for even further speed-up by taking advantage of MATLAB's built-in multi-threading capabilities, as explored here, or by using parallel MATLAB \cite{pMATLAB} or MATLAB's GPU support, future directions for our group.

It is important to point out that a similar or even greater speed-up of the PANDA algorithm should be be obtainable using a lower-level language, including C/C++. The reason the M-code is faster than the current C++ implementation is because we wrote the M-code in terms of matrix operations in order to take advantage of MATLAB's built in support for the BLAS routines which call the underlying SIMD units in the processor. We could have similarity re-implemented PANDA in terms of matrix operations in C++, but that would have required redoing all the data structures; it was far easier to code the algorithm in MATLAB to take advantage of these hardware features. In addition, we suggest that the readability we obtained using the MATLAB/Octave programming language will have a profoundly positive effect on our research abilities. In biological applications we often encounter a wide diversity of data-sets that require thoughtful custom integration in order to leverage them effectively and extract biologically-meaningful results. Future additions and modifications of PANDA considering this information will be far easier using the MATLAB/Octave programming language compared to C++.

In summary, we suggest that the enhanced readability coupled with the significant decrease in time-costs associated with the M-code implementation of PANDA demonstrates a powerful use of the MATLAB/Octave programming language in enabling biomedical research.


\section*{Acknowledgment}

The authors would like to thank Daniel Johnson for an early thoughtful discussion that helped motivate this work.



%

\end{document}